\documentclass[12pt]{emulateapj}
\newcommand{\beq}{\begin{equation}}
\newcommand{\eeq}{\end{equation}}

\newcommand{\hi}{H{\sc i}~}

\newcommand{\citei}[1]{\citeauthor{#1} \citeyear{#1}}

\newcommand{\citeia}[2]{\citeauthor{#1}~(\citeyear{#1};~#2)}
\usepackage[colorlinks,urlcolor=blue,citecolor=blue,linkcolor=blue]{hyperref}
\usepackage{color}
\usepackage{amsmath}
\slugcomment{}

\begin{document}
\title{Ultraviolet Extinction at High Galactic Latitudes II: The Ultraviolet Extinction Function}

\author{J.~E.~G.~Peek\altaffilmark{1}\altaffilmark{2}}
\altaffiltext{1}{Department of Astronomy, Columbia University, New York, NY. jegpeek@gmail.com}
\altaffiltext{2}{Hubble Fellow}

\begin{abstract}
We present a dust-column--dependent extinction curve parameters for ultraviolet wavelengths at high Galactic latitudes. This extinction function diverges from previous work in that it takes into account the results of \citeia{Peek:2013eh}{Paper I}, which demonstrated that there is more reddening in the \emph{GALEX} bands than would be otherwise expected for $E\left(B-V\right) < 0.2$. We also test the biases in the \emph{Planck} and \citeia{SFD98}{SFD} extinction maps, and find that the SFD extinction maps are significantly biased at $E\left(B-V\right) < 0.2$. We find that while an extinction function that that takes into account a varying $R_{FUV}$ with $E\left(B-V\right)$ dramatically improves our estimation of $FUV-NUV$ colors, a fit that also includes \hi column density dependence is superior. The ultraviolet extinction function we present here follows the model of \citet{1999PASP..111...63F}, varying only the amplitude of the $FUV$ rise parameter to be consistent with the data. 
\end{abstract}


\section{Introduction}\label{intro}

In this era of precision extragalactic observations, we must develop equally precise corrections for the effects of Galactic extinction. The Galactic interstellar medium (ISM) is full of small solid particles called dust, which scatter and absorb light across the electromagnetic spectrum, distorting our view of the cosmos (for a review of dust see \citei{2003ARA&A..41..241D}). While the effects of this extinction and reddening are strongest in the Galactic plane, the high latitude sky is especially important to parameterize accurately, as we often examine large groups of extragalactic objects in these areas. 

There are two pieces of data needed to correct for the effects of extinction at high latitude. The first is a map of the overall amplitude of extinction across the sky. Since its publication, this has largely been done using the maps of \citeia{SFD98}{SFD}. SFD uses far infrared (FIR) emission from dust grains supplied by \emph{IRAS}, a temperature correction derived from DIRBE, along with an assumption of a single distribution of grain sizes and compositions to predict the overall dust extinction. Using similar methods, new data from the \emph{Planck} satellite \citep{Collaboration:2013ww} has also been used to create an extinction map (\emph{Planck} Collaboration, \emph{in prep}). The newer \emph{Planck} dust map seems to be more accurate in dusty regions at small scales, owing in part to the fact that the dust temperature correction used is higher resolution than the DIRBE maps employed in SFD (Schlafly \emph{in prep}). The second piece of information needed is the reddening curve (or law); the functional dependence of extinction on wavelength, from the infrared to the ultraviolet. Many reddening laws have been devised, though for Milky Way dust the most often used are \citeia{1989ApJ...345..245C}{CCM}, \citeia{1994ApJ...422..158O}{OD94}, and \citeia{1999PASP..111...63F}{F99}. The essential assumption of this decomposition is that extinction at a given wavelength is linearly dependent upon the dust column density.

The advent of large-area surveys with precision photometry has both necessitated and provided higher precision in our extinction maps. Large photometric surveys, chiefly the Sloan Digital Sky Survey (\citei{york00}; SDSS), but also the \emph{Galaxy Evolution Explorer} (\emph{GALEX}) All-Sky Survey \citeia{2005ApJ...619L...1M}{AIS} and the \emph{Wide-field Infrared Explorer} (\emph{WISE}) All-Sky Data Release \citep{2010AJ....140.1868W}, have allowed the construction of databases of ``standard crayons'' --- objects whose intrinsic colors can be calibrated, such the foreground extinction can be probed. \citeia{2010ApJ...719..415P}{PG10} used quiescent galaxies to probe spatial errors in the SFD maps, while \citeia{2011ApJ...737..103S}{SF11} used spectroscopically observed stars to test reddening laws, and look for overall scaling errors in SFD. \citeia{Jones:2011tf}{JWF11} used M-dwarfs to probe extinction in three dimensions, as well as examine variations in the reddening parameter $R_V$. While reddening laws have always been calibrated against stars at much lower latitudes ($A_V > 1$), these investigations have generally shown that at in the optical wavelengths these reddening laws are rather accurate even at low extinctions. Specifically, SF11 showed that at intermediate latitude the F99 reddening law is more consistent with the data than CCM or OD94 laws. While some variation in the ratio of extinction to reddening, $R_V \equiv A_V/E\left(B-V\right)$, has been seen (e.g. JWF11), the standard assumption of $R_V = 3.1$ has been shown to be a rather good assumption (e.g. in SF11).

In  \citeia{Peek:2013eh}{Paper I} we used galaxies selected from \emph{WISE} and \emph{GALEX} data to investigate extinction in the ultraviolet at high Galactic latitudes, and found large, significant discrepancy with the standard UV reddening laws at high Galactic latitude. Specifically, Paper I showed that extinction in the $NUV$ band was somewhat higher than expected, and the extinction in the $FUV$ was much higher than expected, and that neither of these could be accounted for with any standard extinction curve, using any value of $R_V$. This discrepancy was demonstrated for $E_{\rm SFD} < 0.2$, or about 2/3 of the sky. Paper I also reported significant variations in $FUV-NUV$ color across the high latitude sky in large regions. In this work we extend the analysis of Paper I to build an extinction-dependent reddening law, or ``reddening function'', consistent with observations and as minimally modified from existing laws. To do this accurately we also compare the SFD extinction map to the \emph{Planck} extinction map, and investigate which is less biased at high latitudes, and thus which is a better calibration for an extinction curve. We use the standard nomenclature to describe the total dust column using its optical reddening, $E\left(B-V\right)$, but it may simpler for the reader to interpret this value as the extinction measure $E\left(B-V\right) = A_V/3.1$. We refer to the estimation of each of these values by the two FIR maps as $E_{\rm SFD}$ and $E_{Planck}$.

This paper is laid out as follows. In \S \ref{data} we describe the data sets used in our analysis.  In \S \ref{ebvplanck} we investigate the biases in SFD and \emph{Planck} extinction maps using the PG10 galaxy sample, and confirm our results in \S \ref{nuvplanck} with the \emph{GALEX}-\emph{WISE} sample from Paper I. We confirm the variation of $R_{FUV}$ seen in Paper I using $E_{Planck}$ in \S \ref{fuvnuvred}, and determine its dependency on \hi column in \S \ref{hidep}. We parmeterize the ultraviolet extinction function in \S \ref{UVxf}. We discuss the implication of our results in \S \ref{discussion} and conclude in \S \ref{conclusions}

\section{Data sets}\label{data}

In this work we rely on two types of data sets. The first type consists of maps of foreground extinction, which predict the overall extinction in a given direction of sky. In particular we rely on the SFD map and \emph{Planck} map, each of which directly report measures of $E\left(B-V\right)$ as inferred from emission in the FIR. We also rely on the \hi map from the Leiden-Argentina-Bonn (LAB) survey \citep{Kalberla05}. This is a velocity resolved, stray-radiation corrected map of the Galactic \hi sky in the 21-cm line of neutral hydrogen at 36$^\prime$ resolution. While it is much lower resolution than the FIR maps, it has been shown that combining \hi and FIR maps is often a better predictor of extinction at high Galactic latitude (\citei{2013ApJ...766L...6P}; Peek13). We use this map to construct an \hi column density map of the sky, integrating along Galactic velocities. 

The second type of data set consists of collections of distant galaxies that act as color standards, or standard crayons. The first of these is the PG10 data set of 151,637 galaxies selected from the SDSS main galaxy sample. These galaxies are selected to have neither [O II] and nor H$\alpha$ emission, and thus are not expected to be forming stars. They have very simple and predictable colors and, after correction for color-magnitude, color-redshift, and color-density relations, have typical scatter of only 0.023 magnitudes in $g-r$. The second color standards set is the 373,303 galaxies selected from the \emph{GALEX} and \emph{WISE} data sets, described in Paper I . These galaxies are separated from stars and quasars on the basis of their \emph{WISE} colors, and thus are essentially insensitive to distortion from the effects of reddening (see Paper I figure 1). They are also required to have \emph{GALEX} $NUV < 20$, although can have non-detections in $FUV$. Because of these non-detections, we typically use median $FUV-NUV$ colors when inspecting the colors of collections of these galaxies. We note that any analysis of the variation of the colors of these galaxies needs to take into account the population reddening effect, described in Paper I in \S 4.1 and equation 3, which describes the variation in color of the sample as dim galaxies are extinguished out of it.

\section{Analysis}\label{analysis}

\subsection{Bias in SFD and Planck maps at high Galactic latitudes}\label{ebvplanck}

It is not the aim of this work to provide a final and broad judgement on the relative value of the SFD and \emph{Planck} extinction maps. Not only does such a review go beyond the scope of this work, but it is not at all clear that a single map would be preferable in all contexts. Furthermore, with large new data sets from \emph{Planck}, \emph{WISE} and AKARI \citep{2007PASJ...59S.369M} in the infrared, as well as GASS \citep{McC-G09} and EBHIS \citep{Winkel10} in neutral hydrogen, a new, higher resolution map incorporating all these data will likely soon supersede either map we discuss in this work. In this analysis we limit ourselves to a discussion of the bias in each data set at high latitudes, roughly bounded by $E\left(B-V\right) < 0.2$, where Paper I found significant discrepancy in UV extinction, inconsistent with the literature values. Any bias in this area of sky will modify the overall scaling of any reddening curve we construct, and thus is a critical to investigate and constrain.

To do this analysis, we compare the median color observed in the PG10 standard crayon objects, largely towards the northern Galactic cap, to what is predicted in each map. We use here the $g-r$ colors of the galaxies, as it is the most precise color from PG10 in the estimation of $E\left(B-V\right)$. PG10 showed there were significant discrepancies, up to 45 mmags, in patches of the northern Galactic cap area of the SFD map, though most regions showed deviations of order a few to 10 mmags. In this work, we are instead concerned with the bias of the errors in the reddening maps as a function of $E\left(B-V\right)$: any such trends will manifest themselves as errors in $R_X \equiv A_X/E\left(B-V\right)$. 

\begin{figure*}

\includegraphics[scale=0.95]{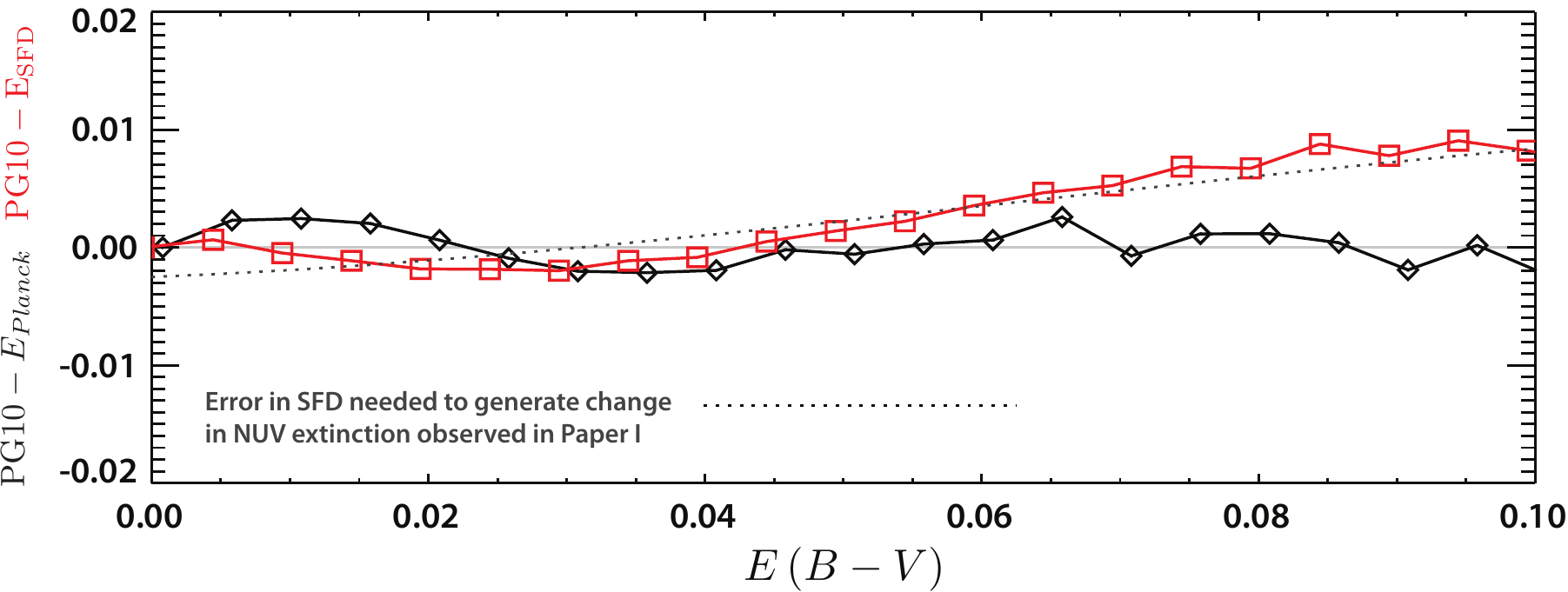}
\caption{We compare the biases of the SFD and \emph{Planck} extinction maps at high Galactic latitude over the SDSS DR7 spectroscopic footprint. In red we show the residual colors of the PG10 galaxies after correcting for Galactic extinction using the SFD map; in black after correcting with the \emph{Planck} map. The x-axis for each residual curve is the \emph{other} extinction map to avoid inducing covariance between the ordinate and abscissa. We show the required bias to reproduce the detected change in $R_{NUV}$ seen in Paper I with a dotted line.}
\label{plancksfd}

\end{figure*}

In Figure \ref{plancksfd} we show the median color of the PG10 galaxies, after having corrected for each of the extinction maps, against overall $E\left(B-V\right)$. We only examine the range $E\left(B-V\right) < 0.1$, as the PG10 area is quite incomplete toward higher extinction. Since we are examining the reddening in a relatively low signal-to-noise area of the maps, we must avoid significant covariance between the independent and dependent variables in our analysis -- such covariance can induce spurious correlations. To do this we chart the residual colors of the PG10 galaxies after correcting with the SFD map against the \emph{Planck} map and vice-versa. 

We find that the SFD-corrected galaxies have a distinct color trend over the range $0.02 < E\left(B-V\right) < 0.1$ that is absent in the \emph{Planck}-corrected galaxies. This under-prediction of $E\left(B-V\right)$ is consistent with a variation in $E_{\rm SFD}$ needed to induce the variation in $R_{NUV}$ seen in Paper I (dotted line in Figure \ref{plancksfd}). We note that SF11 found the opposite trend: SFD \emph{overpredicts} extinction by $\sim15$\%, but this result largely applies to  $E\left(B-V\right) > 0.2$, and thus is not in contradiction to our finding. Indeed, the standardized spectroscopic stars of SF11 show a similar trend over the range $0.02 < E\left(B-V\right) < 0.2$ as we found above. We find consistent results when charting $E_{\rm SFD}$ and $E_{Planck}$ against \hi column density. This result is consistent with the recent work by Liszt (2013), who showed that the observed value of $N\left(\hi\right)/E\left(B-V\right)$ for $E\left(B-V\right) < 0.1$ is in significant excess of the standard canonical value when using $E_{\rm SFD}$. The excess found in that work is consistent with the bias in $E_{\rm SFD}$ we find here.

\subsection{NUV extinction with Planck}\label{nuvplanck}

\begin{figure}

\includegraphics[scale=1.0]{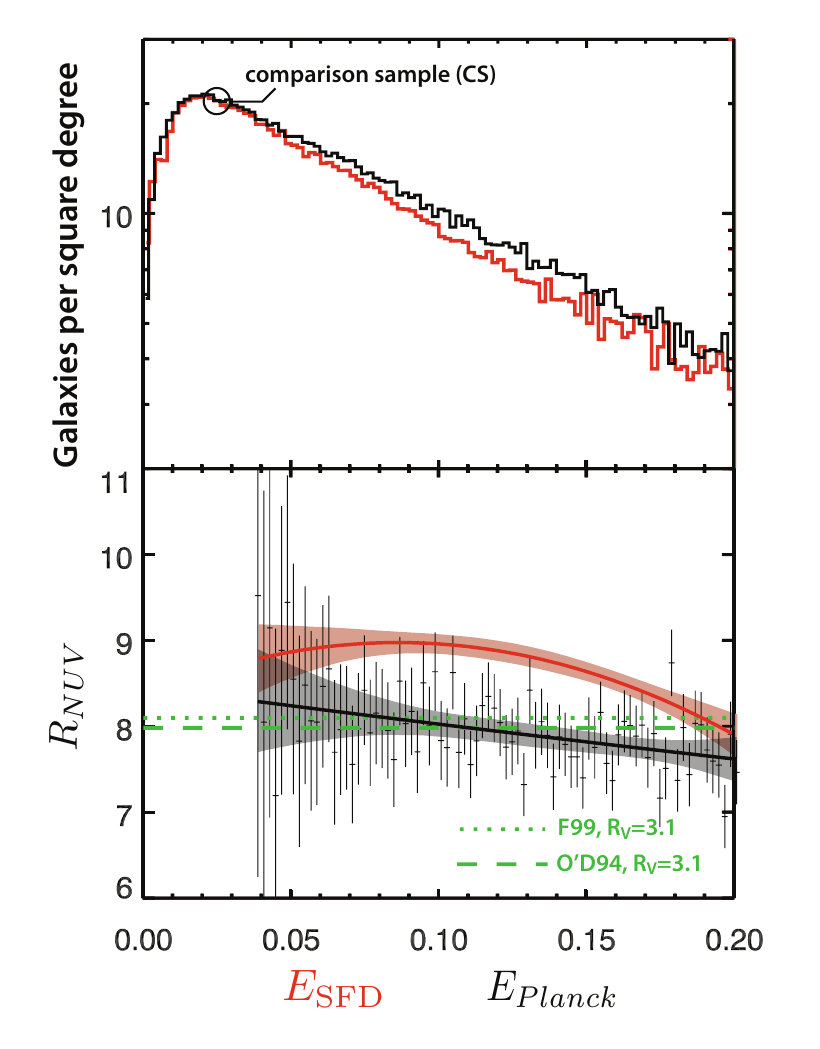}
\caption{In the top panel, the number density of \emph{GALEX}-\emph{WISE} galaxies is shown as function of extinction, for both \emph{Planck} (black) and SFD (red) extinction maps. Over this range galaxies are extinguished out of our sample faster when binned by the SFD data set. The bottom panel shows a quantitative assessment of that statement: binned data and errors show an estimate of $R_{NUV}$ against $E_{Planck}$, and the black line and gray contours show a second-order polynomial fit to the data and the 95\% confidence interval. The same is shown for the SFD data in red (with data points suppressed for clarity). We find no strong trend in  $R_{NUV}$ against $E_{Planck}$ as was found for  $R_{NUV}$ against $E_{SFD}$ in Paper I. }
\label{plancksfd_NUV}
\end{figure}

We note that the SDSS DR7 spectroscopic footprint covers very little sky $E\left(B-V\right) > 0.1$, while the data used to generate the UV extinctions measured in Paper I extend over the entire AIS, a much larger area of sky that includes the Galactic southern hemisphere, and most of the sky where $E\left(B-V\right) < 0.2$. As a further check we replicate the Paper I analysis of the $NUV$ extinction, using the \emph{Planck} map. Simply put, we take a comparison sample of galaxies at where $0.02 < E\left(B-V\right) < 0.03$ and measure their number density. We then artificially extinct the galaxies, removing galaxies too dim to meet the $NUV < 20$ selection criterion, and match the number density of galaxies observed as a function of $E\left(B-V\right)$. We then fit the resulting $R_{NUV}$ with a second-order polynomial, and find confidence intervals. The details of this method can be found in Paper I, \S 4.1; the results are shown in Figure \ref{plancksfd_NUV}. This seems to confirm the result found in the optical: the bulk of the trend detected in the variation in $R_{NUV}$ is an effect of biases in $E_{\rm SFD}$, rather than a change in grain properties at high Galactic latitude. We marginally detect a much weaker trend in $R_{NUV}$, although this trend could easily be generated by small biases in the \emph{Planck} map for $0.1 < E\left(B-V\right) < 0.2$. Since these results are largely consistent with the predictions from the O'D94, CCM and F99 reddening laws, we make the conservative assumption that the reddening laws are accurate in these regions of low extinction in $NUV$.

\begin{figure}
\includegraphics[scale=0.45]{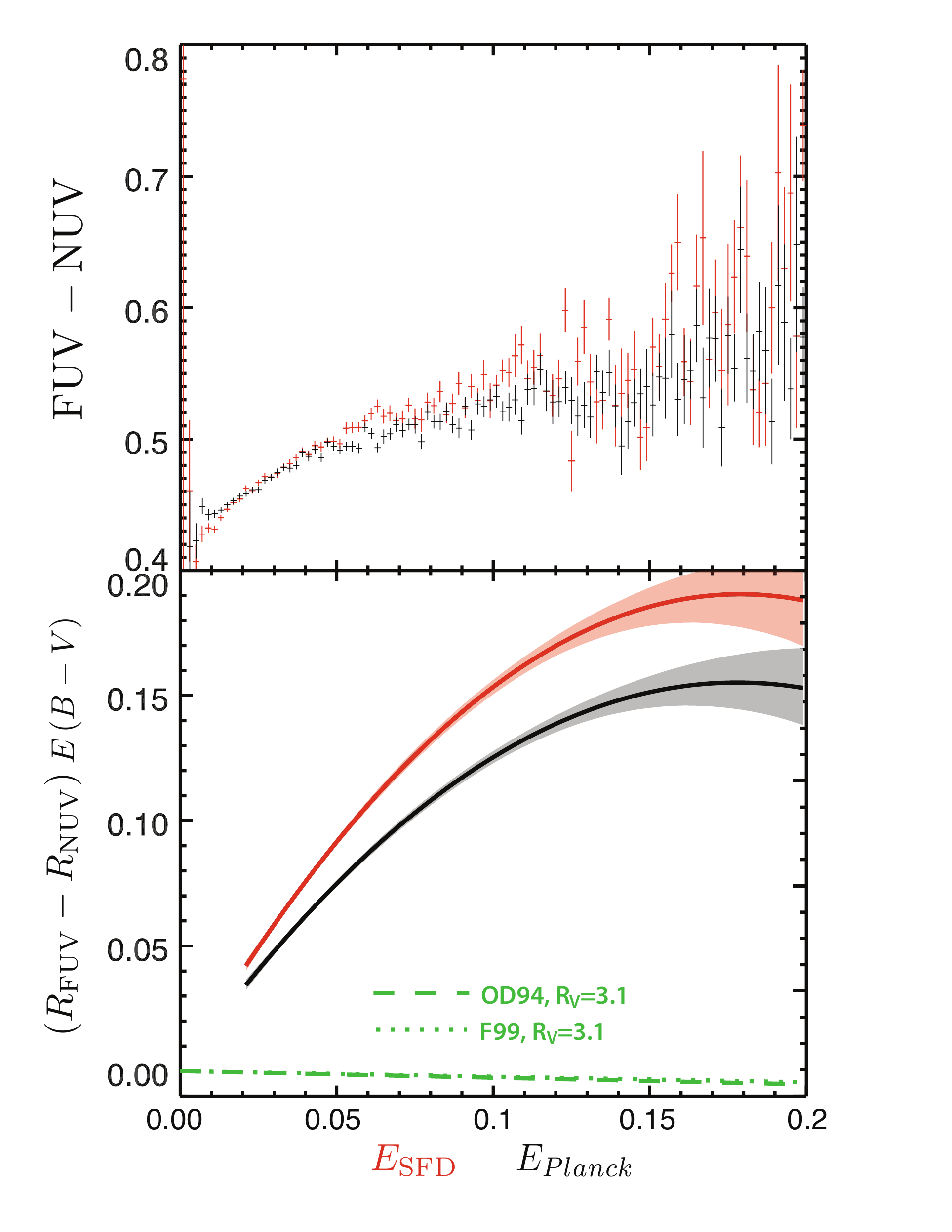}
\caption{The change in the median $FUV-NUV$ color of background galaxies with $E\left(B-V\right)$, for both \emph{Planck} (black) and SFD (red) extinction maps. The top panel shows the median color of \emph{GALEX}-\emph{WISE} galaxies in bins of 0.002 in $E\left(B-V\right)$, with error bars. The bottom panel shows the fit value of $\left(R_{FUV} - R_{NUV}\right)E\left(B-V\right)$ for each extinction map. Also shown is the expected trends from CCM and F99 reddening curves, very clearly inconsistent with the UV reddening trends seen against either extinction map. }
\label{uvcolor}
\end{figure}

\subsection{FUV-NUV reddening with Planck}\label{fuvnuvred}

We also reexamine the the detection of variation in $R_{FUV} - R_{NUV}$ with $E\left(B-V\right)$ found in Paper I. The details of the method are discussed in Paper I \S 4.1, but in essence we find the median color of the \emph{GALEX}-\emph{WISE} galaxies binned by $E\left(B-V\right)$, and fit the resulting data with a second-order polynomial. One important subtlety is that we must compensate for color variation in the underlying population as we go to higher extinction regions and dim objects leave the sample, an effect we called population reddening in Paper I. When we repeat this method using the \emph{Planck} map as our extinction estimator we find \emph{very similar} results to Paper I (Figure \ref{uvcolor}). Indeed, the results are essentially identical, except for the subtle rescaling of the $E\left(B-V\right)$ axis. Since the standard expectation from extinction studies previous to Paper I is that $R_{FUV} - R_{NUV} \simeq 0$, no value of $E\left(B-V\right)$ should produce any variation in the galaxy color. Thus, any rescaling of $E\left(B-V\right)$ cannot induce the observed color shift. Fitting for the population reddening-corrected color of the galaxies is equivalent to fitting for $R_{FUV} - R_{NUV}$ (see equation 5 in Paper I), and we find a best fit value of 

\begin{align}\label{fuvnuv}
&R_{FUV} - R_{NUV} =  \nonumber \\
&\left(1.72 \pm 0.07\right) + \left(-4.75 \pm 0.48 \right) E_{Planck}.
\end{align}

\subsection{The dependence of FUV-NUV reddening on HI}\label{hidep}

\hi has long been known to be a good proxy for dust extinction at high latitude, as originally codified by \citet{Burstein:1978bk}. It was demonstrated more recently in Peek13 that at high Galactic latitudes a combination of FIR and \hi extinction prescriptions are significantly superior to either alone. In light of this fact we explore how \hi can help predict $FUV$ extinction. Our median binning of UV color against $E_{Planck}$ used in \S \ref{fuvnuvred} is not a sufficient method to explore the dependence on two parameters simultaneously, as $N_{\rm HI}$ and $E_{Planck}$ are not allowed to vary independently. Instead, we separate the \emph{GALEX} AIS sky into 16 deg$^2$ regions in a zenith equal area projection, as described in Paper I. In each region we sample the median $E_{Planck}$, \hi column at the galaxy positions, and the median $FUV-NUV$ color, which we call $\tilde{E}_{Planck}$, $\tilde{N}_{\rm HI}$, and $\tilde{C}_{FUV-NUV}$, respectively. We determine the error in the median UV color using a bootstrap analysis, to properly take into account the effects of $FUV$ non-detections. We then fit $\tilde{C}_{FUV-NUV}$ (corrected for population reddening), with a first-order polynomial in $\tilde{N}_{\rm HI}$ and $\tilde{E}_{Planck}$, for a total of 4 parameters, accounting for the $N_{\rm HI}\times E_{Planck}$ cross term. As in \S \ref{fuvnuvred} the overall, unextinguished color of the galaxies is a nuisance parameter, leaving us with a 3 parameter fit for $R_{FUV} - R_{NUV}$:

\begin{align}\label{fuvnuvHI}
&R_{FUV} - R_{NUV} =  \nonumber \\
&\left(-1.01 \pm 0.003\right) + \left( 2.69 \pm 0.12 \right) \frac{N_{\rm HI}}{6.2 \times 10^{21} \rm cm^{-2}}E_{Planck}^{-1} +\nonumber \\
&\left( -4.03 \pm 0.87 \right)\frac{N_{\rm HI}}{6.2 \times 10^{21} \rm cm^{-2}}.
\end{align}

We have normalized the \hi column here by the typical ratio of $N_{\rm HI}/E\left(B-V\right) = 6.2 \times 10^{21} \rm cm^{-2}$ we find for $E_{Planck} < 0.1$, thus this equation is very similar to Equation \ref{fuvnuv} for typical values of $N_{\rm HI}/E_{Planck}$. We can assess the effectiveness of this fit, and the importance of a dependency on \hi, by measuring a $\chi^2$ per degree of freedom (dog). Just using the mean $R_{FUV} - R_{NUV}$ for the whole $E_{Planck} < 0.2$ sky, with no dependence on $N_{\rm HI}$ or $E\left(B-V\right)$, yields $\chi^2/$dof = 16.6: clearly a bad fit to the data. To compare an $E_{Planck}$-only fit to Equation \ref{fuvnuvHI} on even footing, we fit $\tilde{C}_{FUV-NUV}$ with a third-order polynomial in $\tilde{E}_{Planck}$, which yields $\chi^2/$dof = 5.38. The fit in Equation \ref{fuvnuvHI} yields $\chi^2/$dof = 4.55. Thus, while none of these fits fully encompass the variation of $R_{FUV} - R_{NUV}$ across the high-latitude sky, a method making use of both the \hi and FIR maps does the best job. 

\section{Results: The UV Extinction Function}\label{UVxf}

We have shown in \S \ref{analysis} that while the $NUV$ extinction at high Galactic latitudes is largely consistent with the literature when using the \emph{Planck} reddening map, there remains a strong and significant variation in $R_{FUV} - R_{NUV}$, very similar to that shown in Paper I, albeit reduced in amplitude by $\sim$20\%. To capture this variation in a reddening law, we must have a reddening curve that varies with $E\left(B-V\right)$. This is equivalent to saying that at low values of $E\left(B-V\right)$, $FUV$ extinction has a \emph{non-linear dependence} on $E\left(B-V\right)$, and thus is not simply described by a single curve. To highlight this distinction we call our parameterization of extinction in the UV an ``extinction function'', rather than an extinction curve. We note here that we have not shown $E_{Planck}$ to be bias-free in the regime $0.1 < E_{Planck} < 0.2$, such that some of the non-linear dependence on reddening could be due to further biases in $E_{Planck}$. This being said, the steep trends in color we see at lower extinction are clearly real and cannot be due to biases in $E_{Planck}$.

We construct an ultraviolet extinction function that is minimally modified from existing curves, but that is consistent with the observed variation we see in $R_{FUV}$. We base our approach here on the extinction parameterization developed in \citet{Fitzpatrick:1988ef}, and discussed further in \citet{1990ApJS...72..163F} and F99. There are two advantages of this particular parameterization of the extinction. First, it is the basis of the F99 extinction curve shown by SF11 to be a better fit in the optical to the high latitude sky than the CCM / OD94 reddening parameterization. Second, it is separated into six variables, such that we can leave the extinction curve as untouched as possible at wavelengths where the observed extinction is consistent with the curve and modify it only in the $FUV$ region. In this formulation the UV extinction curve is written as

\beq
k\left(\lambda - V\right) \equiv \frac{E\left(\lambda - V\right)}{E\left(B - V\right)}.
\eeq

Using the standard notation of $x \equiv \lambda^{-1}$, the complete parameterization is written as

\beq\label{f99}
k\left(x - V\right) = c_1 + c_2 x + c_3 D\left(x; \gamma, x_0\right) + c_4 F\left(x\right)
\eeq

where $D$ is the Drude profile

\beq
D\left(x; \gamma, x_0\right) = \frac{x^2}{\left(x^2 -x_0^2\right)^2 + x^2\gamma^2}
\eeq

and $F$ is the far-UV rise term

\beq
F\left(x\right)=\begin{cases}
    0.5392\left(x	-5.9\right)^2+0.05644\left(x - 5.9\right)^3, \\
    \text{for $x \geq 5.9 ~\mu m^{-1}$}.\\
    0, \text{for $x < 5.9~ \mu m^{-1}$}.
  \end{cases}
\eeq

\begin{figure}
\includegraphics[scale=1.1]{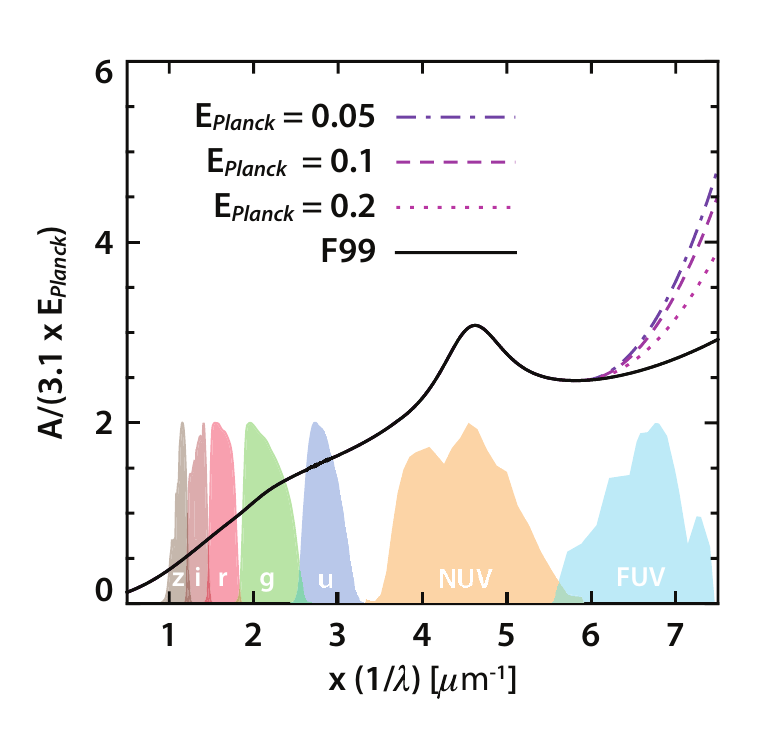}
\caption{The extinction function from Equation \ref{c4ebv}. The standard F99 extinction curve is shown in a solid back lines, with our extinction function shown for discrete values of $E_{Planck}$ in dashed pink, magenta, and purple lines. The transmission functions for SDSS (left) and \emph{GALEX} (right) are shown for comparison. In this Figure we represent the extinction function as dependent on only $E_{Planck}$  (Equation \ref{c4ebv}), as it is easier to visualize than the function dependent on both $E_{Planck}$ and \hi (Equation \ref{c4ebvhi}) }
\label{extfunc}
\end{figure}

Thus, the UV extinction curve is parameterized by the six coefficients $c_1, c_2, c_3, c_4, x_0,$ and $\gamma$. F99 gives specific values for these coefficients; $c_1$ and $c_2$ as functions of the observed $R_V$, and the other four fixed, for typical sight lines. We base our extinction curve on these values with the assumption of $R_V = 3.1$. Since we do not have evidence for the modification of the $NUV$ extinction, we leave $c_3$, $x_0$, and $\gamma$ fixed, as they apply only to the amplitude, position, and width of the 2175 \AA  ~bump, which occurs only in the $NUV$ \emph{GALEX} channel (see Figure \ref{extfunc}). $c_1$ and $c_2$ modify the extinction in both the $NUV$ and $FUV$ channels, and there does indeed exist a family of values of these parameters that leave $NUV$ extinction fixed while modifying $FUV$ extinction. While we cannot rule this modification out, we find it unlikely that $c_1$ and $c_2$ would dramatically depart from their observed relationship with $R_V$ at high latitudes and also conspire to leave $NUV$ extinction fixed. This then leaves us with the single variable parameter $c_4$, the amplitude of the far-UV rise term $F\left(x\right)$. 

To accurately determine the dependence of $c_4$ on $E\left(B - V\right)$ at high latitude, we use the sample of 4000 UV spectral energy distributions from Paper I that are consistent with the selection criteria of the \emph{GALEX}-\emph{WISE} galaxy sample. We then weight the \emph{GALEX} $NUV$ and $FUV$ transmission function by the average of these SEDs to determine the relationship between $FUV-NUV$ reddening in our sample and the parameters of the extinction curve. Based on our results in Equation \ref{fuvnuv}, we find

\beq\label{c4ebv}
c_4 = 4.64  -11.66~E_{Planck}.
\eeq

Alternatively, we can use the results from Equation \ref{fuvnuvHI}, incorporating the \hi dependence:

\begin{align}\label{c4ebvhi}
&c_4 = 1.45 + 6.58\left( \frac{N_{\rm HI}}{6.2 \times 10^{21}~ \rm cm^{-2}}\right) E_{Planck}^{-1} \nonumber \\
& - 9.9\left(\frac{N_{\rm HI}}{6.2 \times 10^{21}~ \rm cm^{-2}}\right)
\end{align}

The extinction function is shown in Figure \ref{extfunc}. We chart four representative extinction curves, for different values of $E_{Planck}$, using the relation from Equation \ref{c4ebv}.

\section{Discussion}\label{discussion}

The first result from this work in \S \ref{ebvplanck} and corroborated in \S \ref{nuvplanck} is that $R_{NUV}$ does not significantly depart from the expected value, when measured against $E_{Planck}$. The parameterization of $R_{NUV}$ discussed in Paper I still holds against $E_{\rm SFD}$, but we no longer expect $E_{\rm SFD}$ to be an unbiased measure of extinction at $E_{\rm SFD} < 0.2$. Conveniently, none of the qualitative conclusions of Paper I depend on a physical interpretation of a varying $R_{NUV}$. It is worth pointing out that we since we see only an overall increase in $R_{FUV}$ at high latitude, it is likely that whatever grain type causes this effect only contributes to extinction in the $FUV$. This is consistent with the general sense that $c_4$ is only weakly correlated with the other extinction coefficients at lower latitude (e.g. \citei{Fitzpatrick:1988ef}), and that the FUV-rise stems from a resonance line similar to the 2175 \AA\ bump (\citei{Joblin:1992bu}, \citei{Li:2001gk}).

We note that the extinction functions derived in \S \ref{UVxf} are, at least naively, not consistent with the literature. As an example, \citet{2004ApJ...616..912V} provides extinction curves toward 417 stars, and parameterizes them with the F99 method (Equation \ref{f99}). While these stars all have  $E\left(B - V\right) \ge 0.2$, and are thus not in the parameter range over which we fit our extinction functions, 84 stars have $E\left(B - V\right) < 0.3$ and these stars do not show any clear trend toward higher $c_4$ values consistent with \S \ref{UVxf}. Rather than posit some very sharp turnover at $E\left(B - V\right) = 0.2$ (which is not supported by the few \emph{GALEX}-\emph{WISE} objects we have toward higher extinctions), we suggest that this discrepancy is due to the fact that the Galactic stars are measured at low latitudes, and thus the bulk of the obscuration may stem from denser gas. It is, after all, not the overall column density of the gas that is important, but rather the physical state of the gas, likely largely determined by its volume density and history (e.g. \citei{Wakker:2000ka}, \citei{Fitzpatrick:1996hg}). Thus, we recommend against using the extinction functions provided in \S \ref{UVxf} for Galactic objects unless they are expected to be well into the halo, and rather refer the reader to the standard F99 UV extinction curves. It is unclear whether these equations can be extrapolated to extragalactic objects toward $E\left(B - V\right) > 0.2$; we do not have enough data at higher extinctions to constrain these functions in this regime. There is certainly no evidence that one should extrapolate these functions to values where $c_4 < 0.41$, the standard F99 value, which occurs for $E\left(B - V\right) > 0.36$ in Equation \ref{fuvnuv}. We also note that while the standard far-UV rise term, $F\left(x\right)$, is the most reasonable parameterization of the variation in $FUV$ extinction we measure, we have not reported direct evidence for extinction beyond the \emph{GALEX} $FUV$ band, which drops off abruptly shortward of 1350\AA.

Equation \ref{fuvnuvHI} may lend some insight into how grains evolve in the diffuse ISM. As \hi becomes molecular, or so dense and cold that it becomes optically thick, the ratio of \hi to FIR tends to decrease, typically for $E_{Planck} > 0.1$ (\citei{Reach:1994hu}, \citei{Douglas:2007kj}). The large, positive coefficient to the $N_{\rm HI}/E_{Planck}$ term indicates that there is more relative extinction in the $FUV$ in regimes where \hi is not at higher densities. This is qualitatively consistent with the interpretation that the grains responsible for the Far-UV rise are plentiful in low density environments, and in higher density environments they tend to be destroyed or agglomerate onto larger grains. 

\section{Conclusions}\label{conclusions}

We present in this work a new ultraviolet extinction function at high Galactic latitude. To summarize our findings:

\begin{enumerate}
\item{While the dependence on $R_{NUV}$ on $E_{\rm SFD}$ found in Paper I holds, we find that $E_{\rm SFD}$ is a biased measure of $E\left(B-V\right)$ for $E\left(B-V\right) < 0.1$, while $E_{Planck}$ is relatively bias-free in this regime.}
\item{$R_{NUV}$ derived against $E_{Planck}$ is indeed largely consistent with the canonical results from the literature.}
\item{$R_{FUV} - R_{NUV}$ when measured against $E_{Planck}$ is still found to be in gross inconsistency with the the literature values, and is qualitatively consistent with what was found in Paper I.}
\item{A non-linear fit in $E_{Planck}$ to $R_{FUV} - R_{NUV}$ can reproduce much of the variation seen in the data (Equation \ref{fuvnuv}), though a similar fit incorporating a dependence on $N_{\rm HI}$ is superior (Equation \ref{fuvnuvHI}).}
\item{A standard F99 extinction curve can reproduce the variation we see in $R_{FUV} - R_{NUV}$ if the $c_4$ parameter is allowed to vary with $E_{Planck}$ (Equation \ref{c4ebv}), or with both $E_{Planck}$ and $N_{\rm HI}$ (Equation \ref{c4ebvhi}).}
\end{enumerate}

To fully understand the variation and origin of the FUV-rise at high Galactic latitude spectroscopic observations of halo stars are needed. Without this spectroscopic information we do not know whether the UV color variation shown in Paper I and this work can truly be encompassed by strong variation in the $c_4$ parameter or are wholly different in character. 

\acknowledgements

This publication makes use of data products from the Wide-field Infrared Survey Explorer, which is a joint project of the University of California, Los Angeles, and the Jet Propulsion Laboratory/California Institute of Technology, funded by the National Aeronautics and Space Administration. {\it Galaxy Evolution Explorer} is a NASA Small Explorer, launched in April 2003. We gratefully acknowledge NASA's support for construction, operation, and science analysis for the \emph{GALEX} mission, developed in cooperation with the Centre National d'Etudes Spatiales of France and the Korean Ministry of Science and Technology. 

This paper was based (in part) on observations obtained with \emph{Planck} (http://www.esa.int/Planck), an ESA science mission with instruments and contributions directly funded by ESA Member States, NASA, and Canada.

Funding for the Sloan Digital Sky Survey (SDSS) and SDSS-II has been provided by the Alfred P. Sloan Foundation, the Participating Institutions, the National Science Foundation, the U.S. Department of Energy, the National Aeronautics and Space Administration, the Japanese Monbukagakusho, and the Max Planck Society, and the Higher Education Funding Council for England. The SDSS Web site is http://www.sdss.org/.

The SDSS is managed by the Astrophysical Research Consortium (ARC) for the Participating Institutions. The Participating Institutions are the American Museum of Natural History, Astrophysical Institute Potsdam, University of Basel, University of Cambridge, Case Western Reserve University, The University of Chicago, Drexel University, Fermilab, the Institute for Advanced Study, the Japan Participation Group, The Johns Hopkins University, the Joint Institute for Nuclear Astrophysics, the Kavli Institute for Particle Astrophysics and Cosmology, the Korean Scientist Group, the Chinese Academy of Sciences (LAMOST), Los Alamos National Laboratory, the Max-Planck-Institute for Astronomy (MPIA), the Max-Planck-Institute for Astrophysics (MPA), New Mexico State University, Ohio State University, University of Pittsburgh, University of Portsmouth, Princeton University, the United States Naval Observatory, and the University of Washington.

JEGP was supported by HST-HF-51295.01A, provided by NASA through a Hubble Fellowship grant from STScI, which is operated by AURA under NASA contract NAS5-26555. JEGP thanks David Schiminovich, Karin Sandstrom, Eddie Schalfly, and Doug Finkbeiner for many helpful suggestions.

\bibliographystyle{yahapj}
\expandafter\ifx\csname natexlab\endcsname\relax\def\natexlab#1{#1}\fi


\end{document}